%% file: manuscriptarxiv.tex
\documentclass[aps,twocolumn,superscriptaddress,showpacs,amsmath,amssymb]{revtex4-1}
\usepackage{subfigure}

\usepackage{relsize}
\def\babar{\mbox{\slshape B\kern-0.1em{\smaller A}\kern-0.1em
    B\kern-0.1em{\smaller A\kern-0.2em R}}}

\usepackage{graphicx} % Include figure files
\usepackage{dcolumn}  % Align table columns on decimal point
\usepackage{comment}

\usepackage{amssymb}
\usepackage{amsmath}
\usepackage{appendix}
\usepackage{color}

\graphicspath{{ps}}

\newcommand{\bb}{b\bar{b}}

\newcommand{\ryns}{R_{\Upsilon(n{\rm S})\pi\pi}}

\newcommand{\ry}{R_{\Upsilon(n{\rm S})\pi\pi}}

\newcommand{\borne}{\sigma^0_{\mu\mu}}

\newcommand{\mevmass}{\text{MeV}/c^2}

\newcommand{\ar}{A_\mathrm{c}}
\newcommand{\anr}{A_\mathrm{ic}}

\begin{document}

\vspace*{-3\baselineskip}
\resizebox{!}{3cm}{\includegraphics{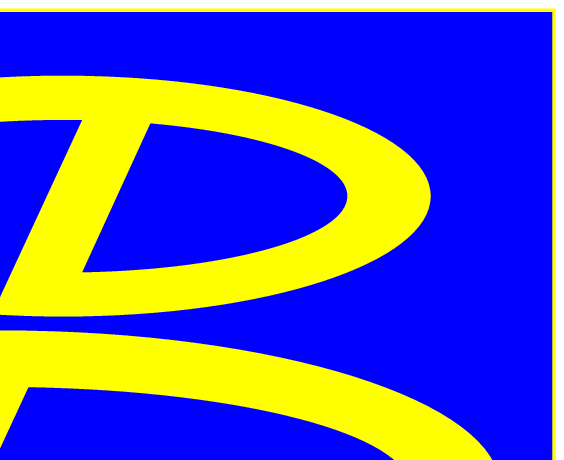}}

%\preprint{\vbox{ \hbox{   }
%						\hbox{Belle DRAFT {\it YY-NN}}
%                        \hbox{Intended for {\it Physical Review Letters}}
%                        \hbox{Author: D. Santel, K. Kinoshita, Y.-P. Yang, P. Chang}
%                        \hbox{Committee: K. Trabelsi(chair),}
%                        \hbox{S. Uozumi, X. Wang, A. Drutskoy}
  		              % \hbox{hep-ex nnnn}
%}}

\title{ \quad\\[1.0cm] 
Measurements of the $\Upsilon$(10860) and $\Upsilon$(11020) resonances via {\boldmath $\sigma(e^+ e^- \to \Upsilon(n{\rm S})\pi^+\pi^-)$
}}

%%%% >>>>> insert the authorlist here. BEFORE the abstract !!!!! <<<<<
%%%% >>>>> from the authorship confirmation web page
%%% Name the file author.tex and use \input{author} to insert into your latex file.
\input{author}

\begin{abstract}
We report new measurements of the total cross sections for $e^+e^-\to \Upsilon(n{\rm S})\pi^+\pi^-$ ($n$ = 1, 2, 3) and $e^+e^-\to b\bar b$ from a high-luminosity fine scan of the region $\sqrt{s} = 10.63$-$11.05$ GeV with the Belle detector.
We observe that the $\Upsilon(n{\rm S})\pi^+\pi^-$ spectra have little or no non-resonant component and  extract from them the masses and widths of  $\Upsilon(10860)$ and $\Upsilon(11020)$ and their relative phase.
We find $M_{10860}=(10891.1\pm3.2^{+0.6}_{-1.7})$ MeV/$c^2$ and $\Gamma_{10860}=(53.7^{+7.1}_{-5.6}\,^{+1.3}_{-5.4})$ MeV and report first measurements $M_{11020}=(10987.5^{+6.4}_{-2.5}\,^{+9.0}_{-2.1})$ MeV/$c^2$, $\Gamma_{11020}=(61^{+9}_{-19}\,^{+2}_{-20})$ MeV, and $\phi_{\rm 11020}-\phi_{\rm 10860} = (-1.0\pm0.4\,^{+1.4}_{-0.1})$ rad.

\end{abstract}

\pacs{13.25.Gv, 14.40.Pq}

\maketitle

%%%% >>>> keep the final version single-spaced
%\tighten

{\renewcommand{\thefootnote}{\fnsymbol{footnote}}}
\setcounter{footnote}{0}

%\tableofcontents

%\section{Background}
The $\Upsilon$(10860)~\cite{ups5sdiscovery,ups6sdiscovery} has historically been interpreted to consist dominantly of the $\Upsilon$(5S), the radial excitation of the S-wave spin-triplet $b\bar{b}$ bound state with $J^{PC}=1^{--}$. 
However, there have been questions about its nature since shortly after its discovery, due to its unexpectedly high mass~\cite{PRD29,PRD34}. 
The Belle Collaboration has observed unexpected behavior in events containing bottomonia among $e^+e^-$ annihilation events at and near the $\Upsilon$(10860).
The rate for $e^+e^-\rightarrow\Upsilon(n{\rm S})\pi^+\pi^-$ ($n$ = 1, 2, 3) at the $\Upsilon$(10860) peak (center of mass energy $\sqrt{s}=10.866\pm0.002$~GeV) is two orders of magnitude larger than that for $\Upsilon(n {\rm S})\rightarrow\Upsilon(1{\rm S})\pi^+\pi^-$ ($n$~=~2, 3, 4)~\cite{chen}. 
Rates to $h_b(m {\rm P})\pi^+\pi^-$ ($m$ = 1, 2) are of the same order of magnitude as to $\Upsilon(n {\rm S})\pi^+\pi^-$, despite the $\Upsilon(5{\rm S})\to h_b(m{\rm P})\pi^+\pi^-$ process requiring a $b$-quark spin-flip~\cite{hbpaper}.
An analysis of $\Upsilon(n{\rm S})\pi^+\pi^-$ ($n$~=~1, 2, 3) and $h_b(m{\rm P})\pi^+\pi^-$ ($m$~=~1, 2) reveals a rich structure, with large contributions from two new bottomonium-like resonance candidates $Z_b(10610)^\pm$ and $Z_b(10650)^\pm$~\cite{garmash}.
%These findings are not easily reconciled with an interpretation of $\Upsilon(10860)$ as a pure $\Upsilon$(5S) state.
Also suggestive is the finding that the peak of $\ry\equiv\sigma(\Upsilon(n{\rm S})\pi^+\pi^-)/\borne$ near $\Upsilon(10860)$ occurs at a mass $9\pm4$~$\mevmass$ higher than that of the $\Upsilon(10860)$, derived from $R_b\equiv\sigma(\bb)/\borne$~\cite{scan}.
[$\sigma^0_{\mu^+\mu^-}  =  (4\pi\alpha^2)/3s$  is the Born $e^+e^-\rightarrow\mu^+\mu^-$ cross-section, with $\alpha$ being the fine-structure constant.]
Here we report on new measurements of $\ry$ and $R_b$, made with a large number of additional scan points between 10.60 and 11.05 GeV. 
$\Upsilon(10860)$ and $\Upsilon(11020)$ will be abbreviated as ``$\Upsilon(5{\rm S})$'' and ``$\Upsilon(6{\rm S})$,'' respectively, for the remainder of this article.

The data were recorded with the Belle detector~\cite{belledetector} at the KEKB~\cite{kekb1,*kekb2} $e^+e^-$ collider.
The Belle detector is a large-solid-angle magnetic spectrometer that consists of a silicon vertex detector (SVD), a 50-layer central drift chamber (CDC), an array of aerogel threshold Cherenkov counters (ACC), a barrel-like arrangement of time-of-flight scintillation counters (TOF), and an electromagnetic calorimeter comprised of CsI(Tl) crystals (ECL), all located inside a superconducting solenoid coil that provides a 1.5~T magnetic field.  An iron flux-return located outside of the coil is instrumented to detect $K_L^0$ mesons and to identify muons (KLM).%  The detector is described in detail elsewhere~\cite{belledetector}.

The data consist of 121.4~fb$^{-1}$ from three energy points very near the $\Upsilon(5{\rm S})$ peak ($\sqrt{s}=10.866\pm0.002$~GeV); approximately $1$~fb$^{-1}$ at each of the six energy points  above 10.80~GeV, studied in Ref.~\cite{scan};  $1$~fb$^{-1}$ at each of 16 new points between $10.63$ and $11.02$~GeV; and $50$~pb$^{-1}$ at each of 61 points taken in 5~MeV steps between $10.75$ and $11.05$~GeV.
For each energy point the data will be categorized as PEAK (on-resonance), HILUM ($\int{\cal L}\sim 1$~fb$^{-1}$) or LOLUM ($\int{\cal L}\sim 50$~pb$^{-1}$).
We measure $\ry$ at the 16 new HILUM sets as well as the six previous HILUM sets and three PEAK sets.
We measure $R_b$ in each of the 61 LOLUM  sets and in the 16 new HILUM sets.
The non-resonant $q\bar q$ continuum $(q\in\{u,d,s,c\})$ background is obtained using a 1.03~fb$^{-1}$ data sample taken below the $B\bar B$ threshold, at $\sqrt{s_{\rm ct}} \equiv 10.520$~GeV (where ct denotes the continuum point). This ``$q\bar{q}$ continuum'' background is distinct from the non-resonant $b\bar{b}$ continuum signal that might be present in our data.

The collision center-of-mass (CMS) energy is calibrated in the PEAK set via the $\Upsilon(n{\rm S})\pi^+\pi^- \{\Upsilon(n{\rm S})\to\mu^+\mu^-\}$ ($n=1,2,3$) event sample.
For these events, the resolution on the mass difference $\Delta M\equiv M(\mu\mu\pi\pi)-M(\mu\mu)$ is dominated by the resolution on the momenta of the two pions, which is narrow due to their relatively low momenta.
The world-average $\Upsilon(n{\rm S})$ masses~\cite{pdg} are used to arrive at the CMS energy with an uncertainty of ($\pm0.2(stat)\pm0.5(sys)$)MeV over the three $\Upsilon$ states for each of the three PEAK sets.
The remaining data sets are calibrated using dimuon mass  in $e^+e^- \rightarrow \mu^+\mu^-$ events. 
The peak value of $M^0_{\mu\mu}$ differs from $\sqrt{s}$, primarily due to initial state radiation (ISR).
The difference is determined via Monte Carlo simulation based on the \textsc{kk2f} generator~\cite{kk2f} and fitted to a straight line at 13 values of $\sqrt s$ between 10.75 and 11.05~GeV.
A constant correction is set by requiring that the  $\Upsilon(1{\rm S})\pi^+\pi^-$ and $\mu$-pair calibrations match for the PEAK set.
The systematic uncertainty from this correction on the $\mu$-pair calibrations is 1.0~MeV.
The statistical uncertainties on $\sqrt{s}$ are shown in the supplemental tables\cite{supplemental}.

% BEGIN YNSPIPI STUFF
Candidate $\Upsilon(n{\rm S})[\to \mu^+\mu^-]\pi^+\pi^-$ events are required to have exactly four charged tracks satisfying track quality criteria, with distances of closest approach to the interaction point (IP) of less than 1 cm and 5 cm in the transverse and longitudinal directions, respectively, and with $p_T>100$ MeV/$c$, including two oppositely charged tracks with an invariant mass above $8\,\textrm{GeV}/c^2$, each consistent with the muon and inconsistent with the kaon hypothesis and two oppositely charged tracks, each consistent with the pion and inconsistent with the electron hypothesis. 
Radiative muon pair events with photon conversions, $e^+e^-\to \gamma\mu^+\mu^-[\gamma\to e^+e^-]$, are suppressed by requiring the $\mu^+\mu^-$ and $\pi^+\pi^-$-candidate vertices be separated in the plane transverse to the $e^+$ beam by less than 3 (4.5) mm for $\Upsilon(1{\rm S},2{\rm S})$ ($\Upsilon(3{\rm S})$). 
We require $|M(\mu^+\mu^-\pi^+\pi^-)-\sqrt{s}_i/c^2|<200$ $\mevmass$, where $i$ denotes the data set and the resolution is $\approx 60~\mevmass$. 
Signal candidates are selected by requiring $\delta\Delta M\equiv|\Delta M - (\sqrt{s_i}/c^2-m_{\Upsilon(n\rm S)})|<25$ $\mevmass$, where the $\Delta M$ resolution is $\approx 7$~MeV/$c^2$.
We select sideband events in the range  $50$ $\mevmass$ $<|\delta\Delta M|<100$ $\mevmass$ to estimate background.

%\section{Cross-section calculation}
Reconstruction efficiencies are estimated via MC simulation. 
Because the relative contributions of intermediate resonances such as the $Z^\pm_b$ may vary with $\sqrt s$, the efficiency is modeled analytically as a function of  $s_1\equiv M^2(\Upsilon\pi^+)$, $s_2\equiv M^2(\Upsilon\pi^-)$, and $\sqrt{s}$ using MC datasets generated at six values of $\sqrt{s}$, with the $\sqrt{s}$-dependence of the efficiency parameters modeled by second-order polynomials.
The {\color{black} efficiencies are 42.5-44.5\%, 31-41\%, and 15-35\% over the range of $\sqrt{s}$ for $\Upsilon$(1S), $\Upsilon$(2S), and $\Upsilon$(3S), respectively.}
%\subsection{$\Upsilon\pi\pi$ parameters fixed to $R_b$ values, etc}
Candidates are summed event-by-event after correcting for reconstruction efficiency for each of the signal and sideband samples.
The net signal $N_{\Upsilon(n{\rm S})\pi\pi,i}$ is equal to the signal sum minus half the sideband sum.  
We then evaluate $R_{\Upsilon(n{\rm S})\pi\pi,i}$=$N_{\Upsilon(n{\rm S})\pi^+\pi^-,i}/({\cal L}_i{\cal B}(\Upsilon(n{\rm S})\to \mu^+\mu^-)\sigma^0_{\mu\mu}(\sqrt s_i))$.

The distributions and fits are shown in Figure~\ref{fig:ypipi}.
Previous results for $\Upsilon(5{\rm S})$ and $\Upsilon(6{\rm S})$ have been based on measurements of $R_b$, where the fitted form is a coherent sum of two $S$-wave Breit-Wigner amplitudes and a constant (continuum), plus an incoherent constant:
\begin{equation}
\label{eq:rbmodel}
\begin{array}{ccl}
{\cal F}({\sqrt{s}})&=&|\anr|^2+|\ar+A_{5{\rm S}}e^{i\phi_{5{\rm S}}}f_{5{\rm S}}({\sqrt{s}})\\
&&+A_{6{\rm S}}e^{i\phi_{6{\rm S}}}f_{6{\rm S}}({\sqrt{s}})|^2,
\end{array}
\end{equation}
where $f_{n{\rm S}}=M_{n{\rm S}}\Gamma_{n{\rm S}}/[(s-M_{n{\rm S}}^2)+iM_{n\rm S}\Gamma_{n{\rm S}}]$ and $\ar$ and $\anr$ are coherent and incoherent continuum terms, respectively. 
For $\ryns$ we adapt this function to accommodate possible differences in resonance substructure between the $\Upsilon(5{\rm S})$ and $\Upsilon(6{\rm S})$ and the phase space volume of $\Upsilon(n{\rm S})\pi^+\pi^-$ near the  mass threshold.
$\ar$ and $\anr$ are found to be consistent with, and are thus fixed to, zero in all three channels.
Assuming the resonance substructures are {\it not} identical, the relative phase between the respective (normalized) amplitudes, ${\cal D}_{5S,n}(s_1, s_2)$ and ${\cal D}_{6S,n}(s_1, s_2)$, varies over the Dalitz space $(s_1,s_2)$.
The cross term between the two resonances from Eq.~(\ref{eq:rbmodel}) is
\begin{equation}
\label{eq:crossterm}
\begin{array}{ccl}
2k_n A_{5{\rm S},n}A_{6{\rm S},n}\Re [e^{i\delta_n}f_{5{\rm S}}f^*_{6{\rm S}}],
\end{array}
\end{equation}
where $k_n e^{i\delta_n}\equiv \int{{\cal D}_{5S,n}(s_1, s_2){\cal D}^*_{6S,n}(s_1, s_2)ds_1ds_2}$ and the {\it decoherence coefficient} $k_n$ is in the range $0<k_n<1$.
If the resonance substructures are identical, $k_n$ is unity and $\delta_n\equiv\phi_{5{\rm S}}-\phi_{6{\rm S}}$.
Given the rich structure found at $\sqrt{s}=10.866$ GeV~\cite{garmash}, some deviation of both $k_n$ and $\delta_n$ from this scenario are likely. 
To account for near-threshold behavior, the fitting function is multiplied by $\Phi_n(\sqrt{s})$, the ratio of phase-space volumes of $e^+e^-\rightarrow\Upsilon(n{\rm S})\pi\pi$ to $e^+e^-\rightarrow\Upsilon(n{\rm S})\gamma\gamma$.
The fit function is thus
\begin{equation}
\label{eq:rbmodeldecoherence}
\begin{array}{ccl}
{\cal F}_n^\prime(\sqrt{s}) &=&\Phi_n(\sqrt{s})\cdot\{|A_{5{\rm S},n}f_{5{\rm S}}|^2
+|A_{6{\rm S},n}f_{6{\rm S}}|^2\\
&&+ 2k_n A_{5{\rm S},n}A_{6{\rm S},n}\Re [e^{i\delta_n}f_{5{\rm S}}
f^*_{6{\rm S}}]\}.
\end{array}
\end{equation}
In fitting $\ryns$, the $\Upsilon(5{\rm S})$ and $\Upsilon(6{\rm S})$ masses, widths, and relative phases are allowed to float, constrained to the same values for the three channels.
Due to limited statistics, floating the three $k_n$ and $\delta_n$ did not produce a stable fit, so we 
allow the three $k_n$ to float and constrain the three $\delta_n$ to a common value. 
We find $k_1=1.04\pm 0.19$, $k_2=0.87\pm 0.17$, $k_3=1.07\pm 0.23$, and $\delta_n=-1.0\pm 0.4$.
The results of the fit are shown in Table~\ref{tab:rbfit} and Fig.~\ref{fig:ypipi}.
As a systematic check, we fit with $k_n$ fixed to unity and the three $\delta_n$ allowed to float independently; we find $\delta_1=-0.5 \pm 1.9$, $\delta_2=-1.1 \pm 0.5$, and $\delta_3=1.0 ^{+0.8}_{-0.5}$, while the resonance masses and widths change very little.

To measure $R_b$, we select $b\bar b$ events by requiring at least five charged tracks with transverse momentum $p_T >100$ MeV/$c$ that satisfy track quality criteria based on their impact parameters relative to the IP.
Each event must have more than one ECL cluster with energy above $100$~MeV, a total energy in the ECL between $0.1$ and $0.8\times\sqrt{s}$, and an energy sum of all charged tracks and photons exceeding $0.5\times\sqrt{s}$.
We demand that the reconstructed event vertex be within 1.5 and 3.5~cm of the IP in the transverse and longitudinal dimensions (perpendicular and parallel to the $e^+$ beam), respectively. 
To suppress events of non-$b\bar{b}$ origin, events are further required to satisfy  $R_2<0.2$, where $R_2$ is the ratio of the second and zeroth
Fox-Wolfram moments~\cite{foxwolfram}. 

The selection efficiency $\epsilon_{b\bar b,i}$  for the $i^{\rm th}$ scan set is estimated via MC simulation  based on EvtGen~\cite{evtgen} and GEANT3~\cite{geant}. 
Efficiencies are determined for each type of open $b\bar b$ event found at $\sqrt{s}=10.866$ GeV: $B^{(*)}\bar{B}^{(*)}(\pi)$ and  $B_s^{(*)}\bar{B}_s^{(*)}$.
As the relative rates of the different event types are only known at the on-resonance point, we take the average of the highest and  lowest efficiencies as $\epsilon_{b\bar b}$ and the difference divided by $\sqrt{12}$ as its uncertainty.
The value of  $\epsilon_{b\bar b}$ increases approximately linearly from about $70\%$ to $74\%$ over the scan region. 
The value  at the on-resonance point is in good agreement with $\epsilon_{b\bar b}$ determined with the known event mixture~\cite{pdg}.

{\color{black}
Events passing the above criteria include direct $b\bar b$, $q\bar q$   continuum ($q=u,d,s,c$), and bottomonia produced via ISR: $e^+e^-\to \gamma\Upsilon(n{\rm S})$ ($n$=1, 2, 3).
The number of selected events is 
%\begin{eqnarray}
\begin{equation}
%\scriptstyle
%\small
{N_{i}}=  {\mathcal{L}_i} \times\left[\sigma_{b\bar b,i}\epsilon_{b\bar b,i}+\sigma_{q\bar{q},i}\epsilon_{q\bar{q},i}+\sum\sigma_{\mathrm{ISR},i}\epsilon_{\mathrm{ISR},i}\right]
\label{eq:ns}
\end{equation}
where $\mathcal L_i$ is the integrated luminosity of data set $i$ and the sum is over the three $\Upsilon$ states produced via ISR.
The contribution from $\sigma(q\bar q)$, which scales as $1/s$, is estimated from the data taken at $\sqrt{s_{\rm ct}}$, where $\sigma_{b\bar b}=0$, and is corrected for luminosity and energy differences.
The subtracted quantity  
%terms, $R^0_b(s)$ and $R_{b,\rm{ISR}}(s)$:   
\begin{equation}
\tilde R_{b,i}=\frac{1}{\epsilon_{b\bar b}}\left(\frac{N_{i}}{\mathcal{L}_i\sigma^0_{\mu\mu,i}}-\frac{N_{\rm ct}}{\mathcal{L}_{\rm ct}\sigma^0_{\mu\mu,{\rm ct}}}\frac{\epsilon_{q\bar{q},i}}{\epsilon_{q\bar{q},{\rm ct}}}\right)
\label{eq:rb0}
\end{equation}
includes a residual contribution from ISR, which differs from $q\bar q$ continuum in its $s$-dependence.
For comparison with a previous measurement by BABAR~\cite{babarrb}, we define $R_b$ to include the ISR events; we use Ref.~\cite{ISR} and measured electronic widths of $\Upsilon(nS)$ to calculate $\sigma_{\rm{ISR}}$.
Although the nature of the $b\bar b$ continuum is not known, it is known that the ISR contribution is not flat in $\sqrt{s}$, so we also calculate $R_{b,i}^\prime\equiv R_{b,i} - \sum\sigma_{{\rm ISR},i}/\sigma^0_{\mu^+\mu^-,i}$. 
These measurements yield the visible cross-sections and include neither corrections due to ISR events containing $\{b\bar{b}\}$ final states above $B\bar{B}$ threshold nor the vacuum polarization necessary to obtain the Born cross-section~\cite{kuraev}.

Both $\{R_{b,i}\}$ and $\{R_{b,i}^\prime\}$ are fitted to ${\cal F}$ (Eq.~\ref{eq:rbmodel});
the fitting range is restricted to $10.82$-$11.05$~GeV to avoid complicated threshold effects below 10.8~GeV~\cite{tornqvist}.
The resulting masses, widths, and relative phase for $\{R_{b,i}^\prime\}$ are shown in Table~\ref{tab:rbfit}; they do not differ significantly between $\{R_{b,i}\}$ and $\{R_{b,i}^\prime\}$.
Those for $R_b$ are consistent with those from earlier measurements by Belle~\cite{scan} and BABAR~\cite{babarrb}.
The $R_b^\prime$ data and fit are shown in Fig.~\ref{fig:ypipi}. 

That the $\Upsilon(n{\rm S})\pi^+\pi^-$ occurs only in resonance events in the $\Upsilon(5{\rm S})$ region, i.e., the continuum components $\ar$ and $\anr$ are consistent with zero, is in marked contrast to the large resonance-continuum interference reflected in the $R_b^\prime$ fit. 
The relationship of the various $b\bar b$ final states to the resonance and continuum may help to elucidate the nature of the resonance and of $b\bar b$ hadronization in this complex threshold region.
As a first probe, we evaluate the rates at  $\sqrt{s}=10.866$ of $\Upsilon(n{\rm S})\pi^+\pi^-$ and other states known to have essentially no continuum content, to be compared with the resonance rate obtained from $R_b^\prime$.
The ``$\Upsilon(5{\rm S})$ resonance rate'' corresponds to the term  that includes $|f_{5{\rm S}}|^2$ in Eqs.~(\ref{eq:rbmodeldecoherence}) and~(\ref{eq:rbmodel}).
%In principle, the term proportional to the absolute square of the $\Upsilon$(5S) amplitude in $\ryns$, summed with corresponding terms for all other event types, should result in the corresponding term for  $R_b^\prime$, i.e.,
%\begin{eqnarray*}
%\sum_i{\{P_i\equiv|A_{5{\rm S}}(i) f_{5{\rm S}}|^2\times \Phi_i}\}\\
%=\{P_b\equiv|A_{5{\rm S}}(R_b) f_{5{\rm S}}|^2\}
%\end{eqnarray*}
%where the sum is over all resonance final states.
We calculate $P_n\equiv|A_{5{\rm S}}(n{\rm S}) f_{5{\rm S}}|^2\times \Phi_n$ ($n=1,2,3$) and $P_b$ at the on-resonance energy point ($\sqrt{s}=10.866$~GeV) using the results from the fits to $\ryns$ and $R_b^\prime$, respectively.
We find ${\cal P}\equiv\sum_n P_n/P_b$={ $0.170\pm 0.009$}. 
We argue that a number of known related final states measured in the PEAK data are expected to behave similarly, i.e., to contain very little continuum: $\Upsilon(n{\rm S})\pi^0\pi^0$~\cite{z0paper}, which is related by isospin to $\Upsilon(n{\rm S})\pi^+\pi^-$;
$h_b(m{\rm P})\pi^+\pi^-$ ($m=1,\,2$), which is found to be saturated by $Z_{b}^\pm\pi^\mp$~\cite{garmash,hbpaper} a state included in $\Upsilon(n{\rm S})\pi^+\pi^-$; $h_b(m{\rm P})\pi^0\pi^0$, which is expected by isospin symmetry to occur at half the rate of $h_b(m{\rm P})\pi^+\pi^-$.
Assuming isospin symmetry and taking the rate of $h_b(m{\rm P})\pi^+\pi^-$ ($m=1,\,2$) measured in PEAK data,~\cite{hbpaper}, we include these states and obtain ${\cal P}=0.42\pm0.04$.
Another class of states that is likely to be similarly resonance-dominated is $B^*B^{(*)}\pi$~\cite{garmashConf}:
preliminary evidence indicates that $[B^*B^{(*)}]^\pm\pi^\mp$ is consistent with originating exclusively from $Z_b^\pm\pi^\mp$.  
Taking the preliminary measurement and again assuming that isospin symmetry holds for $[B^*B^{(*)}]^0\pi^0$, we find ${\cal P}=1.09\pm0.15$.

A value of ${\cal P}=1$ corresponds to the saturation of the ``5S'' amplitude by the contributing channels.
It is surprising to find ${\cal P}$ so close to unity, as it implies little room in the resonance for other known final states such as $B_{(s)}^{(*)}\bar B_{(s)}^{(*)}$,  which comprise nearly 20\% of $b\bar b$ events at the peak\cite{drutskoybsinclusive}.
More significantly, it is inconsistent with the large resonance-continuum interference found in the fit to $R_b^\prime$ (Fig.~\ref{fig:ypipi}) because the channels contributing to ${\cal P}$ include little or no continuum.
It has long been known that a flat continuum distribution in this complex region that includes many $\{b\bar b\}$ mass thresholds is overly simplistic~\cite{tornqvist}, and we conclude that this internal inconsistency of the $R_b^\prime$ fit, elucidated by ${\cal P}$, is likely due to the model's na{\" i}vet{\' e}. 
This finding leads to the conclusion that masses and widths for the $\Upsilon$(10860) and $\Upsilon$(11020) obtained from $R_b^{(\prime)}$ carry unknown systematic uncertainties due to the unknown shape of the continuum and its interaction with the resonance, which may vary with energy.
The results reported here for the masses, widths, and relative phase of the $\Upsilon$(10860) and $\Upsilon$(11020) are thus from the $\Upsilon(n{\rm S})\pi^+\pi^-$ analysis, which are robust due to low continuum content.

We have considered the following sources of systematic uncertainty: integrated luminosity, event selection efficiency, energy calibration, reconstruction efficiency, secondary branching fractions, and fitting procedure. 
The effects of the uncertainties in $R_b^{(\prime)}$ and $\ryns$ on $\Upsilon(5{\rm S})$ and $\Upsilon(6{\rm S})$ parameters depend on whether they are correlated or not over the data sets at different energy points.  
The overall uncertainty in the integrated luminosity is 1.4\%, while the uncorrelated variation is 0.1$\%$-0.2$\%$.
%The overall uncertainty in $\sqrt{s}$  is 1~MeV.
%The uncertainty in the $b\bar{b}$ event selection efficiency, {\color{black} $\epsilon_{b\bar{b}}$}, stems from uncertainties in the mix of event types, containing $B_q$, $B_s$, bottomonia, tau pairs, two-photon events, and $q\bar{q}$ continuum, is estimated to be 1.1\% (uncorrelated).
The uncertainty in the $b\bar{b}$ event selection efficiency, {\color{black} $\epsilon_{b\bar{b}}$}, stems from uncertainties in the mix of event types, containing $B^{(*)}$, $B_s^{(*)}$, and bottomonia and is estimated to be 1.1\% (uncorrelated).
The uncertainty on $\ryns$ for each $\Upsilon(n{\rm S})$ is dominated by those on the branching fractions, $\mathcal{B}(\Upsilon(n{\rm S})\rightarrow\mu^+\mu^-)$~\cite{pdg}: $\pm$2\%, $\pm$10\%, and $\pm$10\% for $n$~=~1, 2, and 3, respectively. 
%{\color{black} In the fit of $\ryns$, the values of $\phi_{6{\rm S}}$, $M_{6{\rm S}}$ and $\Gamma_{6{\rm S}}$ are fixed to the result from fitting $R_b-R_{\rm{ISR}}$, and their uncertainty is taken as the statistical errors from the same fit. }
{\color{black}The uncertainties from possible non-zero $\ar$ and/or $\anr$ in $\ryns$ are obtained by allowing them to float in the fit and taking the variation of the fitted values of the other parameters with respect to default results.}
Possible biases due to constraints on $k_n$ and $\delta_n$ in the fit are estimated by taking the shifts found by varying the constraints and included as systematic errors.
%The event-by-event efficiency correction to obtain $\ryns$ is insensitive, but not immune, to intermediate states in the three-body decay. 
%MC studies of $\Upsilon(5$S$,6$S$)\to \Upsilon(n$S$)\pi^+\pi^-$ event sets generated at several values of $\sqrt s$ and with models that include both twice the nominal contribution from $Z_b$-like resonances and no $Z_b$ contribution yield uncertainties of $0.11\%$, $0.55\%$ and $0.92\%$ for $n=1$, 2, and 3, respectively.
The lower end of the fit range is varied between 10.63 and 10.82~GeV. 
Approximate radiative corrections to the visible cross-section measurements are made, as in Ref.~\cite{kuraev}, and the fits are repeated. The combined systematic uncertainties and fit results appear in Table~\ref{tab:rbfit}.

To summarize, we have measured the cross sections for $e^+e^-\to \Upsilon(n{\rm S})\pi^+\pi^-$ ($n$~=~1, 2, 3) and $e^+e^-\to b\bar b$  in the region $\sqrt{s}=10.8$-$11.05$~GeV to determine masses and widths for $\Upsilon(10860)$ and $\Upsilon(11020)$.
From $\ry$ we find 
$M_{10860}=10891.1\pm3.2(stat)^{+0.6}_{-1.7}(sys)\pm 1.0(\sqrt{s})$ MeV/$c^2$, 
$\Gamma_{10860}=53.7^{+7.1}_{-5.6}\,^{+1.3}_{-5.4}$ MeV, 
$M_{11020}=10987.5^{+6.4}_{-2.5}(stat)^{+9.0}_{-2.1}(sys)\pm 1.0(\sqrt{s})$ MeV/$c^2$, 
$\Gamma_{11020}=61^{+9}_{-19}\,^{+2}_{-20}$ MeV, and 
$\phi_{11020}-\phi_{10860} = -1.0\pm0.4\,^{+1.4}_{-0.1}$~rad.
We find that $\ry$ is dominated by the two resonances, with $b\bar{b}$ continuum consistent with zero.
Although the resonance masses and widths obtained from $R_b^\prime$ are consistent with those from $\ry$, the validity of using a flat continuum in the $R_b^\prime$ fit is brought into question by incompatibilities between the fitted amplitudes for $\ry$ and $R_b^\prime$.
We thus report only results from $\ry$.
We do not see the peaking structure at 10.9 GeV in the $R_b$ distribution that was suggested by A.~Ali {\it et al.}~\cite{ali2} based on the BABAR measurement of $R_b$~\cite{babarrb}. 
We set an upper limit on $\Gamma_{ee}$  for the proposed structure of 9~eV with a 90\% confidence level.

%The amplitude for the $\Upsilon(10860)$ component of $R_b^\prime$ appears to be saturated by  $\Upsilon(n{\rm S})\pi^+\pi^-$ and several other non-$B_{(s)}^{(*)}$-pair events. Paradoxically, this leaves little room for $B_{(s)}^{(*)}$-pairs and is inconsistent with the large interference between $\Upsilon(10860)$ and $b\bar{b}$ continuum implied from the fit to $R_b^\prime$.

%We have used $\Upsilon$(5S) as a shorthand for $\Upsilon(10860)$, but the nature of this structure remains uncertain.

We thank the KEKB group for excellent operation of the
accelerator; the KEK cryogenics group for efficient solenoid
operations; and the KEK computer group, the NII, and 
PNNL/EMSL for valuable computing and SINET4 network support.  
We acknowledge support from MEXT, JSPS and Nagoya's TLPRC (Japan);
ARC and DIISR (Australia); FWF (Austria); NSFC (China); MSMT (Czechia);
CZF, DFG, and VS (Germany); DST (India); INFN (Italy); 
MOE, MSIP, NRF, GSDC of KISTI, and BK21Plus (Korea);
MNiSW and NCN (Poland); MES (particularly under Contract No. 14.A12.31.0006) and RFAAE (Russia); ARRS (Slovenia);
IKERBASQUE and UPV/EHU (Spain); 
SNSF (Switzerland); NSC and MOE (Taiwan); and DOE and NSF (USA).

\nocite{supplemental}

\begin{figure}[htb]
\includegraphics[width=9.50cm]{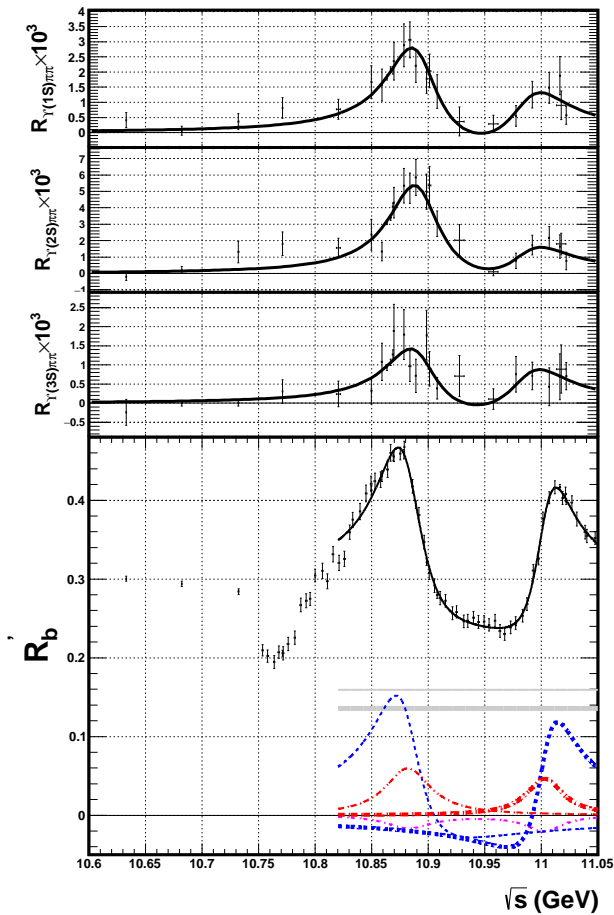}
		\caption{(From top) $\ry$ data with results of our nominal fit for $\Upsilon(1{\rm S})$; $\Upsilon(2{\rm S})$; $\Upsilon(3{\rm S})$; $R_b^\prime$, data with components of fit: total (solid curve), constants $|\anr|^2$ (thin), $|\ar|^2$ (thick); for $\Upsilon(5{\rm S})$ (thin) and 	$\Upsilon(6{\rm S})$ (thick): $|f|^2$ (dot-dot-dash), cross terms with $\ar$ (dashed), and two-resonance cross term (dot-dash).   Error bars include the statistical and uncorrelated systematic uncertainties.}
		\label{fig:ypipi}
\end{figure}

\begin{table*}[htb]
\caption{ $\Upsilon(5{\rm S})$ and $\Upsilon(6{\rm S})$ masses, widths, and phase difference, extracted from fits to data. 
The errors are statistical and systematic. The 1~MeV uncertainty on the  masses due to the systematic uncertainty in $\sqrt{s}$ is not included.}
\label{tab:rbfit}
\begin{tabular}{lcccccc}

\hline \hline
 & $M_{5{\rm S}}$ ($\mevmass$) & $\Gamma_{5{\rm S}}$ (MeV) & $M_{6{\rm S}}$ ($\mevmass$) & $\Gamma_{6{\rm S}}$ (MeV) & $\phi_{6{\rm S}}$-$\phi_{5{\rm S}}$ $(\delta)$ (rad)& $\chi^2/dof$ \\ \hline

%$R_b$ & $10881.9\pm 1.0\pm1.2$ &$49.8\pm1.9\,^{+2.1}_{-2.8}$ & $11002.9\pm 1.1\,^{+0.8}_{-0.9}$ & $38.5^{+1.6}_{-1.5}\,^{+1.3}_{-2.4}$ & $-1.86^{+0.24}_{-0.10}\pm 0.10$ & 55/50\\
$R_b^\prime$  & $10881.8^{+1.0}_{-1.1}\pm1.2$ & $48.5^{+1.9}_{-1.8}\,^{+2.0}_{-2.8}$ &  $11003.0\pm 1.1\,^{+0.9}_{-1.0}$ & $39.3^{+1.7}_{-1.6}\,^{+1.3}_{-2.4}$ & $-1.87^{+0.32}_{-0.51}\pm0.16$  & 56/50\\

$\ryns$ & $10891.1\pm3.2^{+0.6}_{-1.7}$ & $53.7^{+7.1}_{-5.6}\,^{+1.3}_{-5.4}$ & $10987.5^{+6.4}_{-2.5}\,^{+9.0}_{-2.1}$ & $61^{+9}_{-19}\,^{+2}_{-20}$ & $ -1.0\pm0.4\,^{+1.4}_{-0.1}$ & 51/56\\

\hline \hline
\end{tabular}
\end{table*}

\def\urlprefix{}
\def\url#1{}
%\bibliography{references}
\bibliographystyle{apsrev4-1}
\bibliography{references2}
\clearpage
\begin{center}
\input{suppTables.tex}	
\end{center}

\end{document}

%% file: author.tex
%%% Paper:    e+ e- -> Y(nS) pi+ pi-, b bbar
%%% Journal:  Physical Review Letters
%%% Contacts: D. Santel (santeldm@mail.uc.edu)
%%%           K. Kinoshita(kay.kinoshita@uc.edu)
%%%           Y.-P. Yang (yjp1986@hep1.phys.ntu.edu.tw)
%%%           P. Chang(pchang@phys.ntu.edu.tw)
%%% Non-responding authors or those who said NO are commented out.
%%% ====================================================================
%%% Click the RELOAD button on your web browser to see the updated file.
%%% ====================================================================
%%% Use \input{author} to insert this material into your latex file.
%%%%% Force institutions to appear in alphabetical order when typeset.
\noaffiliation
\affiliation{University of the Basque Country UPV/EHU, 48080 Bilbao}
%%%\affiliation{Beihang University, Beijing 100191}
\affiliation{University of Bonn, 53115 Bonn}
\affiliation{Budker Institute of Nuclear Physics SB RAS and Novosibirsk State University, Novosibirsk 630090}
\affiliation{Faculty of Mathematics and Physics, Charles University, 121 16 Prague}
%%%\affiliation{Chiba University, Chiba 263-8522}
%%%\affiliation{Chonnam National University, Kwangju 660-701}
\affiliation{University of Cincinnati, Cincinnati, Ohio 45221}
\affiliation{Deutsches Elektronen--Synchrotron, 22607 Hamburg}
%%%\affiliation{Department of Physics, Fu Jen Catholic University, Taipei 24205}
\affiliation{Justus-Liebig-Universit\"at Gie\ss{}en, 35392 Gie\ss{}en}
%%%\affiliation{Gifu University, Gifu 501-1193}
%%%\affiliation{II. Physikalisches Institut, Georg-August-Universit\"at G\"ottingen, 37073 G\"ottingen}
\affiliation{The Graduate University for Advanced Studies, Hayama 240-0193}
\affiliation{Gyeongsang National University, Chinju 660-701}
\affiliation{Hanyang University, Seoul 133-791}
\affiliation{University of Hawaii, Honolulu, Hawaii 96822}
\affiliation{High Energy Accelerator Research Organization (KEK), Tsukuba 305-0801}
%%%\affiliation{Hiroshima Institute of Technology, Hiroshima 731-5193}
\affiliation{IKERBASQUE, Basque Foundation for Science, 48011 Bilbao}
%%%\affiliation{University of Illinois at Urbana-Champaign, Urbana, Illinois 61801}
%%%\affiliation{Indian Institute of Technology Bhubaneswar, Satya Nagar 751007}
\affiliation{Indian Institute of Technology Guwahati, Assam 781039}
\affiliation{Indian Institute of Technology Madras, Chennai 600036}
%%%\affiliation{Indiana University, Bloomington, Indiana 47408}
\affiliation{Institute of High Energy Physics, Chinese Academy of Sciences, Beijing 100049}
\affiliation{Institute of High Energy Physics, Vienna 1050}
%%%\affiliation{Institute for High Energy Physics, Protvino 142281}
%%%\affiliation{Institute of Mathematical Sciences, Chennai 600113}
\affiliation{INFN - Sezione di Torino, 10125 Torino}
\affiliation{Institute for Theoretical and Experimental Physics, Moscow 117218}
\affiliation{J. Stefan Institute, 1000 Ljubljana}
\affiliation{Kanagawa University, Yokohama 221-8686}
%%%\affiliation{Institut f\"ur Experimentelle Kernphysik, Karlsruher Institut f\"ur Technologie, 76131 Karlsruhe}
%%%\affiliation{Kavli Institute for the Physics and Mathematics of the Universe (WPI), University of Tokyo, Kashiwa 277-8583}
\affiliation{Kennesaw State University, Kennesaw GA 30144}
\affiliation{Department of Physics, Faculty of Science, King Abdulaziz University, Jeddah 21589}
\affiliation{Korea Institute of Science and Technology Information, Daejeon 305-806}
\affiliation{Korea University, Seoul 136-713}
%%%\affiliation{Kyoto University, Kyoto 606-8502}
\affiliation{Kyungpook National University, Daegu 702-701}
\affiliation{\'Ecole Polytechnique F\'ed\'erale de Lausanne (EPFL), Lausanne 1015}
\affiliation{Faculty of Mathematics and Physics, University of Ljubljana, 1000 Ljubljana}
\affiliation{Luther College, Decorah, Iowa 52101}
\affiliation{University of Maribor, 2000 Maribor}
\affiliation{Max-Planck-Institut f\"ur Physik, 80805 M\"unchen}
\affiliation{School of Physics, University of Melbourne, Victoria 3010}
\affiliation{Moscow Physical Engineering Institute, Moscow 115409}
\affiliation{Moscow Institute of Physics and Technology, Moscow Region 141700}
\affiliation{Graduate School of Science, Nagoya University, Nagoya 464-8602}
\affiliation{Kobayashi-Maskawa Institute, Nagoya University, Nagoya 464-8602}
%%%\affiliation{Nara University of Education, Nara 630-8528}
\affiliation{Nara Women's University, Nara 630-8506}
%%%\affiliation{National Central University, Chung-li 32054}
%%%\affiliation{National United University, Miao Li 36003}
\affiliation{Department of Physics, National Taiwan University, Taipei 10617}
\affiliation{H. Niewodniczanski Institute of Nuclear Physics, Krakow 31-342}
%%%\affiliation{Nippon Dental University, Niigata 951-8580}
\affiliation{Niigata University, Niigata 950-2181}
%%%\affiliation{University of Nova Gorica, 5000 Nova Gorica}
\affiliation{Osaka City University, Osaka 558-8585}
%%%\affiliation{Osaka University, Osaka 565-0871}
\affiliation{Pacific Northwest National Laboratory, Richland, Washington 99352}
%%%\affiliation{Panjab University, Chandigarh 160014}
\affiliation{Peking University, Beijing 100871}
\affiliation{University of Pittsburgh, Pittsburgh, Pennsylvania 15260}
%%%\affiliation{Punjab Agricultural University, Ludhiana 141004}
%%%\affiliation{Research Center for Electron Photon Science, Tohoku University, Sendai 980-8578}
%%%\affiliation{Research Center for Nuclear Physics, Osaka University, Osaka 567-0047}
%%%\affiliation{RIKEN BNL Research Center, Upton, New York 11973}
%%%\affiliation{Saga University, Saga 840-8502}
\affiliation{University of Science and Technology of China, Hefei 230026}
\affiliation{Seoul National University, Seoul 151-742}
%%%\affiliation{Shinshu University, Nagano 390-8621}
\affiliation{Soongsil University, Seoul 156-743}
\affiliation{Sungkyunkwan University, Suwon 440-746}
\affiliation{School of Physics, University of Sydney, NSW 2006}
\affiliation{Department of Physics, Faculty of Science, University of Tabuk, Tabuk 71451}
\affiliation{Tata Institute of Fundamental Research, Mumbai 400005}
\affiliation{Excellence Cluster Universe, Technische Universit\"at M\"unchen, 85748 Garching}
\affiliation{Toho University, Funabashi 274-8510}
%%%\affiliation{Tohoku Gakuin University, Tagajo 985-8537}
\affiliation{Tohoku University, Sendai 980-8578}
\affiliation{Department of Physics, University of Tokyo, Tokyo 113-0033}
\affiliation{Tokyo Institute of Technology, Tokyo 152-8550}
%%%\affiliation{Tokyo Metropolitan University, Tokyo 192-0397}
%%%\affiliation{Tokyo University of Agriculture and Technology, Tokyo 184-8588}
%%%\affiliation{University of Torino, 10124 Torino}
%%%\affiliation{Toyama National College of Maritime Technology, Toyama 933-0293}
\affiliation{CNP, Virginia Polytechnic Institute and State University, Blacksburg, Virginia 24061}
\affiliation{Wayne State University, Detroit, Michigan 48202}
\affiliation{Yamagata University, Yamagata 990-8560}
\affiliation{Yonsei University, Seoul 120-749}
  \author{D.~Santel}\affiliation{University of Cincinnati, Cincinnati, Ohio 45221} % Cincinnati
    \author{K.~Kinoshita}\affiliation{University of Cincinnati, Cincinnati, Ohio 45221} % Cincinnati
      \author{P.~Chang}\affiliation{Department of Physics, National Taiwan University, Taipei 10617} % Taiwan
  \author{A.~Abdesselam}\affiliation{Department of Physics, Faculty of Science, University of Tabuk, Tabuk 71451} % Tabuk
  \author{I.~Adachi}\affiliation{High Energy Accelerator Research Organization (KEK), Tsukuba 305-0801}\affiliation{The Graduate University for Advanced Studies, Hayama 240-0193} % KEK
% \author{K.~Adamczyk}\affiliation{H. Niewodniczanski Institute of Nuclear Physics, Krakow 31-342} % Krakow
  \author{H.~Aihara}\affiliation{Department of Physics, University of Tokyo, Tokyo 113-0033} % Tokyo
  \author{S.~Al~Said}\affiliation{Department of Physics, Faculty of Science, University of Tabuk, Tabuk 71451}\affiliation{Department of Physics, Faculty of Science, King Abdulaziz University, Jeddah 21589} % Tabuk
  \author{K.~Arinstein}\affiliation{Budker Institute of Nuclear Physics SB RAS and Novosibirsk State University, Novosibirsk 630090} % BINP
% \author{Y.~Arita}\affiliation{Graduate School of Science, Nagoya University, Nagoya 464-8602} % Nagoya
  \author{D.~M.~Asner}\affiliation{Pacific Northwest National Laboratory, Richland, Washington 99352} % PNNL
% \author{T.~Aso}\affiliation{Toyama National College of Maritime Technology, Toyama 933-0293} % Toyama
% \author{V.~Aulchenko}\affiliation{Budker Institute of Nuclear Physics SB RAS and Novosibirsk State University, Novosibirsk 630090} % BINP
  \author{T.~Aushev}\affiliation{Moscow Institute of Physics and Technology, Moscow Region 141700}\affiliation{Institute for Theoretical and Experimental Physics, Moscow 117218} % ITEP
% ...
% Ljubljana
% \author{N.~Zwahlen}\affiliation{\'Ecole Polytechnique F\'ed\'erale de Lausanne (EPFL), Lausanne 1015} % Lausanne
% \author{O.~Zyukova}\affiliation{Budker Institute of Nuclear Physics SB RAS and Novosibirsk State University, Novosibirsk 630090} % BINP
\collaboration{The Belle Collaboration}

%% file: suppTables.tex
	\begin{table*}
		\begin{center}
			\caption{Ratio $R_{\Upsilon(n{\rm S})\pi\pi}\equiv\sigma(\Upsilon(n{\rm S})\pi^+\pi^-)/\sigma(\mu^+\mu^-)$. Uncertainties on $R_{\Upsilon(n{\rm S})\pi\pi}$ are statistical.}
			\label{tab:xsecratio}
			\scalebox{1}{
				\begin{tabular}{c c c c}
					\hline\hline
					$\sqrt{s}$ (MeV) & $R_{\Upsilon(1{\rm S})\pi\pi}\times10^3$  & $R_{\Upsilon(2{\rm S})\pi\pi}\times10^3$ & $R_{\Upsilon(3{\rm S})\pi\pi}\times10^3$\\ 
					\hline \hline
					$ 10632.8 \pm 0.4\pm 1.0 $ & $ 0.41 \pm 0.25 $ & $ -0.20 \pm 0.22 $ & $ -0.24 \pm 0.34 $\\
					$ 10682.0 \pm 0.4\pm 1.0 $ & $ 0.06 \pm 0.15 $ & $ 0.20 \pm 0.21 $ & $ 0.00 \pm 0.07 $\\
					$ 10732.1 \pm 0.4\pm 1.0 $ & $ 0.36 \pm 0.27 $ & $ 1.31 \pm 0.66 $ & $ 0.00 \pm 0.07 $\\
					$ 10771.1 \pm 0.4\pm 1.0 $ & $ 0.81 \pm 0.34 $ & $ 1.81 \pm 0.71 $ & $ 0.31 \pm 0.31 $\\
					$ 10820.5 \pm 1.8\pm 1.0 $ & $ 0.77 \pm 0.33 $ & $ 1.55 \pm 0.59 $ & $ 0.24 \pm 0.34 $\\
					$ 10849.7 \pm 0.4\pm 1.0 $ & $ 1.67 \pm 0.53 $ & $ 2.36 \pm 0.93 $ & $ 0.32 \pm 0.34 $\\
					$ 10858.9 \pm 0.4\pm 1.0 $ & $ 1.57 \pm 0.52 $ & $ 1.34 \pm 0.57 $ & $ 1.08 \pm 0.49 $\\
					$ 10863.3 \pm 0.2\pm 0.5 $ & $ 1.81 \pm 0.08 $ & $ 3.10 \pm 0.13 $ & $ 0.93 \pm 0.08 $\\
					$ 10866.7 \pm 0.2 \pm 0.5$ & $ 2.08 \pm 0.09 $ & $ 3.40 \pm 0.17 $ & $ 1.07 \pm 0.08 $\\
					$ 10868.6 \pm 0.2\pm 0.5 $ & $ 2.06 \pm 0.12 $ & $ 4.06 \pm 0.23 $ & $ 1.27 \pm 0.12 $\\
					$ 10869.5 \pm 0.4\pm 1.0 $ & $ 2.36 \pm 0.62 $ & $ 4.31 \pm 0.94 $ & $ 1.89 \pm 0.70 $\\
					$ 10878.5 \pm 0.4\pm 1.0 $ & $ 2.88 \pm 0.71 $ & $ 5.35 \pm 1.05 $ & $ 1.79 \pm 0.65 $\\
					$ 10883.6 \pm 0.9\pm 1.0 $ & $ 3.06 \pm 0.60 $ & $ 5.20 \pm 0.92 $ & $ 0.97 \pm 0.40 $\\
					$ 10888.9 \pm 0.4\pm 1.0 $ & $ 2.21 \pm 0.54 $ & $ 5.85 \pm 1.09 $ & $ 0.72 \pm 0.43 $\\
					$ 10898.5 \pm 0.4\pm 1.0 $ & $ 1.78 \pm 0.48 $ & $ 4.98 \pm 1.07 $ & $ 1.78 \pm 0.65 $\\
					$ 10901.1 \pm 1.1\pm 1.0 $ & $ 2.03 \pm 0.55 $ & $ 5.36 \pm 1.15 $ & $ 0.90 \pm 0.45 $\\
					$ 10907.7 \pm 0.4\pm 1.0 $ & $ 1.35 \pm 0.56 $ & $ 3.03 \pm 0.78 $ & $ 0.39 \pm 0.28 $\\
					$ 10927.5 \pm 4.0\pm 1.0 $ & $ 0.37 \pm 0.48 $ & $ 2.03 \pm 0.94 $ & $ 0.71 \pm 0.53 $\\
					$ 10957.5 \pm 4.0\pm 1.0 $ & $ 0.29 \pm 0.29 $ & $ 0.11 \pm 0.25 $ & $ 0.11 \pm 0.27 $\\
					$ 10977.5 \pm 0.4\pm 1.0 $ & $ 0.55 \pm 0.33 $ & $ 0.77 \pm 0.47 $ & $ 0.76 \pm 0.46 $\\
					$ 10991.9 \pm 0.4\pm 1.0 $ & $ 1.25 \pm 0.44 $ & $ 1.72 \pm 0.60 $ & $ 0.71 \pm 0.36 $\\
					$ 11006.8 \pm 0.4\pm 1.0 $ & $ 1.48 \pm 0.51 $ & $ 2.17 \pm 0.70 $ & $ 0.43 \pm 0.49 $\\
					$ 11016.4 \pm 0.4\pm 1.0 $ & $ 1.88 \pm 0.63 $ & $ 1.64 \pm 0.71 $ & $ 0.69 \pm 0.60 $\\
					$ 11017.5 \pm 4.0\pm 1.0 $ & $ 0.90 \pm 0.47 $ & $ 1.79 \pm 0.66 $ & $ 0.89 \pm 0.63 $\\
					$ 11022.0 \pm 0.4\pm 1.0 $ & $ 0.57 \pm 0.30 $ & $ 0.76 \pm 0.54 $ & $ 0.71 \pm 0.36 $\\
					\hline\hline
				\end{tabular}
			}
		\end{center}
	\end{table*}

\begin{table*}
	\begin{center}
		\caption{Ratio $R_{b}^\prime\equiv\sigma(b\bar{b})/\sigma(\mu^+\mu^-)$. Uncertainties on $R_b^\prime$ are statistical, uncorrelated systematic, and correlated systematic, respectively. Uncertainties on $\sqrt{s}$ are statistical; the systematic uncertainty of  $\pm$1.0~MeV is correlated for all points.}
		\label{tab:rbxsecratio1}
		\scalebox{.7}{
			\begin{tabular}{c c}
				\hline\hline
				$\sqrt{s}$ (MeV) & $R_b^\prime$\\
				\hline\hline

				10632.8 $\pm$ 0.4 & 0.3004 $\pm$ 0.0019 $\pm$ 0.0026 $\pm$ 0.0112\\
				10682.0 $\pm$ 0.4 & 0.2940 $\pm$ 0.0018 $\pm$ 0.0029 $\pm$ 0.0117\\
				10732.2 $\pm$ 0.4 & 0.2841 $\pm$ 0.0018 $\pm$ 0.0031 $\pm$ 0.0109\\
				10753.5 $\pm$ 0.6 & 0.2094 $\pm$ 0.0062 $\pm$ 0.0036 $\pm$ 0.0111\\
				10757.9 $\pm$ 0.7 & 0.2022 $\pm$ 0.0067 $\pm$ 0.0041 $\pm$ 0.0110\\
				10763.7 $\pm$ 0.7 & 0.1946 $\pm$ 0.0067 $\pm$ 0.0047 $\pm$ 0.0105\\
				10767.7 $\pm$ 0.7 & 0.2072 $\pm$ 0.0067 $\pm$ 0.0043 $\pm$ 0.0107\\
				10771.1 $\pm$ 0.4 & 0.2056 $\pm$ 0.0017 $\pm$ 0.0034 $\pm$ 0.0109\\
				10771.6 $\pm$ 0.7 & 0.2057 $\pm$ 0.0068 $\pm$ 0.0049 $\pm$ 0.0114\\
				10776.0 $\pm$ 0.7 & 0.2176 $\pm$ 0.0068 $\pm$ 0.0049 $\pm$ 0.0112\\
				10782.0 $\pm$ 0.7 & 0.2257 $\pm$ 0.0068 $\pm$ 0.0049 $\pm$ 0.0112\\
				10787.1 $\pm$ 0.7 & 0.2671 $\pm$ 0.0067 $\pm$ 0.0053 $\pm$ 0.0118\\
				10792.0 $\pm$ 0.7 & 0.2725 $\pm$ 0.0069 $\pm$ 0.0054 $\pm$ 0.0116\\
				10795.5 $\pm$ 0.6 & 0.2748 $\pm$ 0.0063 $\pm$ 0.0052 $\pm$ 0.0111\\
				10799.9 $\pm$ 0.6 & 0.3046 $\pm$ 0.0062 $\pm$ 0.0051 $\pm$ 0.0112\\
				10806.3 $\pm$ 0.7 & 0.3104 $\pm$ 0.0072 $\pm$ 0.0064 $\pm$ 0.0112\\
				10810.7 $\pm$ 0.7 & 0.2972 $\pm$ 0.0072 $\pm$ 0.0060 $\pm$ 0.0112\\
				10815.7 $\pm$ 0.7 & 0.3315 $\pm$ 0.0073 $\pm$ 0.0066 $\pm$ 0.0120\\
				10821.0 $\pm$ 0.7 & 0.3207 $\pm$ 0.0072 $\pm$ 0.0065 $\pm$ 0.0122\\
				10825.9 $\pm$ 0.7 & 0.3256 $\pm$ 0.0072 $\pm$ 0.0063 $\pm$ 0.0113\\
				10830.4 $\pm$ 0.6 & 0.3595 $\pm$ 0.0066 $\pm$ 0.0060 $\pm$ 0.0120\\
				10833.2 $\pm$ 0.7 & 0.3752 $\pm$ 0.0068 $\pm$ 0.0066 $\pm$ 0.0113\\
				10839.6 $\pm$ 0.7 & 0.3858 $\pm$ 0.0075 $\pm$ 0.0070 $\pm$ 0.0116\\
				10845.0 $\pm$ 0.7 & 0.4086 $\pm$ 0.0075 $\pm$ 0.0073 $\pm$ 0.0135\\
				10849.4 $\pm$ 0.7 & 0.4211 $\pm$ 0.0069 $\pm$ 0.0064 $\pm$ 0.0114\\
				10849.7 $\pm$ 0.4 & 0.4127 $\pm$ 0.0018 $\pm$ 0.0056 $\pm$ 0.0115\\
				10852.8 $\pm$ 0.7 & 0.4242 $\pm$ 0.0072 $\pm$ 0.0072 $\pm$ 0.0132\\
				10857.7 $\pm$ 0.7 & 0.4262 $\pm$ 0.0073 $\pm$ 0.0068 $\pm$ 0.0123\\
				10858.9 $\pm$ 0.4 & 0.4307 $\pm$ 0.0018 $\pm$ 0.0064 $\pm$ 0.0117\\
				10863.9 $\pm$ 0.7 & 0.4392 $\pm$ 0.0075 $\pm$ 0.0070 $\pm$ 0.0119\\
				10866.7 $\pm$ 0.7 & 0.4602 $\pm$ 0.0076 $\pm$ 0.0071 $\pm$ 0.0125\\
				10869.0 $\pm$ 0.3 & 0.4562 $\pm$ 0.0015 $\pm$ 0.0060 $\pm$ 0.0103\\
				10869.5 $\pm$ 0.4 & 0.4554 $\pm$ 0.0018 $\pm$ 0.0057 $\pm$ 0.0117\\
				10875.2 $\pm$ 0.5 & 0.4592 $\pm$ 0.0052 $\pm$ 0.0061 $\pm$ 0.0121\\
				10878.5 $\pm$ 0.4 & 0.4586 $\pm$ 0.0018 $\pm$ 0.0056 $\pm$ 0.0117\\
				10878.8 $\pm$ 0.4 & 0.4678 $\pm$ 0.0043 $\pm$ 0.0055 $\pm$ 0.0119\\
				10886.0 $\pm$ 0.6 & 0.4176 $\pm$ 0.0067 $\pm$ 0.0054 $\pm$ 0.0115\\
				10888.9 $\pm$ 0.4 & 0.3943 $\pm$ 0.0018 $\pm$ 0.0043 $\pm$ 0.0113\\
				10891.8 $\pm$ 0.7 & 0.3814 $\pm$ 0.0072 $\pm$ 0.0060 $\pm$ 0.0111\\

				\hline\hline

			\end{tabular}
						\begin{tabular}{c c}
							\hline\hline
							$\sqrt{s}$ (MeV) & $R_b^\prime$\\
							\hline\hline
							10896.2 $\pm$ 0.7 & 0.3469 $\pm$ 0.0072 $\pm$ 0.0049 $\pm$ 0.0134\\
							10898.5 $\pm$ 0.4 & 0.3213 $\pm$ 0.0017 $\pm$ 0.0037 $\pm$ 0.0109\\
							10900.9 $\pm$ 0.7 & 0.3078 $\pm$ 0.0071 $\pm$ 0.0045 $\pm$ 0.0106\\
							10905.6 $\pm$ 0.7 & 0.2917 $\pm$ 0.0070 $\pm$ 0.0051 $\pm$ 0.0104\\
							10907.7 $\pm$ 0.4 & 0.2794 $\pm$ 0.0017 $\pm$ 0.0036 $\pm$ 0.0104\\
							10910.4 $\pm$ 0.7 & 0.2759 $\pm$ 0.0068 $\pm$ 0.0050 $\pm$ 0.0106\\
							10915.2 $\pm$ 0.7 & 0.2724 $\pm$ 0.0068 $\pm$ 0.0040 $\pm$ 0.0120\\
							10921.5 $\pm$ 0.7 & 0.2565 $\pm$ 0.0068 $\pm$ 0.0047 $\pm$ 0.0109\\
							10925.0 $\pm$ 0.7 & 0.2581 $\pm$ 0.0067 $\pm$ 0.0039 $\pm$ 0.0110\\
							10931.3 $\pm$ 0.7 & 0.2463 $\pm$ 0.0067 $\pm$ 0.0046 $\pm$ 0.0107\\
							10934.8 $\pm$ 0.7 & 0.2469 $\pm$ 0.0068 $\pm$ 0.0046 $\pm$ 0.0113\\
							10940.0 $\pm$ 0.7 & 0.2507 $\pm$ 0.0067 $\pm$ 0.0046 $\pm$ 0.0110\\
							10944.4 $\pm$ 0.7 & 0.2456 $\pm$ 0.0067 $\pm$ 0.0044 $\pm$ 0.0112\\
							10949.3 $\pm$ 0.7 & 0.2458 $\pm$ 0.0067 $\pm$ 0.0036 $\pm$ 0.0112\\
							10953.7 $\pm$ 0.7 & 0.2418 $\pm$ 0.0066 $\pm$ 0.0044 $\pm$ 0.0106\\
							10959.0 $\pm$ 0.7 & 0.2470 $\pm$ 0.0067 $\pm$ 0.0045 $\pm$ 0.0122\\
							10963.3 $\pm$ 0.7 & 0.2340 $\pm$ 0.0067 $\pm$ 0.0043 $\pm$ 0.0108\\
							10967.4 $\pm$ 0.7 & 0.2300 $\pm$ 0.0066 $\pm$ 0.0044 $\pm$ 0.0104\\
							10972.7 $\pm$ 0.7 & 0.2385 $\pm$ 0.0066 $\pm$ 0.0044 $\pm$ 0.0105\\
							10977.3 $\pm$ 0.7 & 0.2443 $\pm$ 0.0067 $\pm$ 0.0043 $\pm$ 0.0103\\
							10977.5 $\pm$ 0.4 & 0.2439 $\pm$ 0.0016 $\pm$ 0.0028 $\pm$ 0.0099\\
							10983.3 $\pm$ 0.7 & 0.2533 $\pm$ 0.0067 $\pm$ 0.0044 $\pm$ 0.0102\\
							10987.3 $\pm$ 0.7 & 0.2680 $\pm$ 0.0067 $\pm$ 0.0044 $\pm$ 0.0109\\
							10991.9 $\pm$ 0.4 & 0.2974 $\pm$ 0.0017 $\pm$ 0.0030 $\pm$ 0.0100\\
							10992.7 $\pm$ 0.7 & 0.3107 $\pm$ 0.0066 $\pm$ 0.0042 $\pm$ 0.0116\\
							10997.5 $\pm$ 0.7 & 0.3262 $\pm$ 0.0069 $\pm$ 0.0048 $\pm$ 0.0107\\
							
							11001.3 $\pm$ 0.7 & 0.3766 $\pm$ 0.0071 $\pm$ 0.0048 $\pm$ 0.0105\\
							11006.8 $\pm$ 0.4 & 0.3997 $\pm$ 0.0018 $\pm$ 0.0031 $\pm$ 0.0108\\
							11006.9 $\pm$ 0.7 & 0.4032 $\pm$ 0.0068 $\pm$ 0.0037 $\pm$ 0.0112\\
							11012.1 $\pm$ 0.7 & 0.4165 $\pm$ 0.0073 $\pm$ 0.0041 $\pm$ 0.0113\\
							11016.4 $\pm$ 0.4 & 0.4167 $\pm$ 0.0020 $\pm$ 0.0026 $\pm$ 0.0107\\
							11018.8 $\pm$ 0.7 & 0.4029 $\pm$ 0.0073 $\pm$ 0.0039 $\pm$ 0.0119\\
							11021.4 $\pm$ 0.7 & 0.4080 $\pm$ 0.0073 $\pm$ 0.0050 $\pm$ 0.0107\\
							11022.0 $\pm$ 0.4 & 0.3983 $\pm$ 0.0017 $\pm$ 0.0031 $\pm$ 0.0106\\
							11026.9 $\pm$ 0.7 & 0.3971 $\pm$ 0.0072 $\pm$ 0.0049 $\pm$ 0.0110\\
							11031.3 $\pm$ 0.7 & 0.3762 $\pm$ 0.0072 $\pm$ 0.0048 $\pm$ 0.0109\\
							11038.6 $\pm$ 0.7 & 0.3596 $\pm$ 0.0072 $\pm$ 0.0047 $\pm$ 0.0111\\
							11040.2 $\pm$ 0.7 & 0.3547 $\pm$ 0.0071 $\pm$ 0.0046 $\pm$ 0.0113\\
							11047.4 $\pm$ 0.7 & 0.3521 $\pm$ 0.0072 $\pm$ 0.0037 $\pm$ 0.0110\\
							\hline\hline

						\end{tabular}
		}
	\end{center}
\end{table*}